\documentclass[useAMS,usenatbib]{mnras}
\usepackage{graphicx}
\usepackage{textcomp}
\usepackage{amssymb}\usepackage{hyperref}

 \usepackage{times}

\newcommand\lsim{\mathrel{\rlap{\lower4pt\hbox{\hskip1pt$\sim$}}
        \raise1pt\hbox{$<$}}}
\newcommand\gsim{\mathrel{\rlap{\lower4pt\hbox{\hskip1pt$\sim$}}
        \raise1pt\hbox{$>$}}}
\newcommand{\be}{\begin{equation}}
\newcommand{\ba}{\begin{eqnarray}}
\newcommand{\ee}{\end{equation}}
\newcommand{\ea}{\end{eqnarray}}

\title[White dwarf resonant detonation during binary inspiral]{On the resonant detonation of sub-Chandrasekhar mass white dwarfs during binary inspiral}

\author[B. McKernan, K.E.S.Ford]{B. McKernan$^{1,2,3,4}$\thanks{E-mail:bmckernan at amnh.org, sford at amnh.org}, K.E.S. Ford$^{1,2,3,4}$\\
$^{1}$Department of Science, BMCC, City University of New York, New York, NY 10007, USA\\ 
$^{2}$Department of Astrophysics, American Museum of Natural History, New York, NY 10024, USA\\ 
$^{3}$Graduate Center, City University of New York, 365 5th Avenue, New York, NY 10016, USA\\
$^{4}$Kavli Institute for Theoretical Physics, UC Santa Barbara, CA 93106, USA\\
}

\begin{document}

\date{Accepted. Received; in original form}

\pagerange{\pageref{firstpage}--\pageref{lastpage}} \pubyear{2008}

\maketitle

\label{firstpage}

\begin{abstract}
White dwarfs (WDs) are believed to detonate via explosive Carbon-fusion in a Type Ia Supernova when their temperature and/or density reach the point where Carbon is ignited in a runaway reaction. Observations of the Type Ia supernova (SN) rate imply all WD binaries that merge through the emission of gravitational radiation within a Hubble time should result in SNe, regardless of total mass. Here we investigate the conditions under which a single WD in a binary system might extract energy from its orbit, depositing enough energy into a resonant mode such that it detonates before merger. We show that, ignoring non-linear effects, in a WD binary in tidal lock at small binary separations, the sustained tidal forcing of a low-order quadrupolar g-mode or a harmonic of a  low-order quadrupolar p-mode could in principle drive the average temperature of Carbon nuclei in the mode over the runaway fusion threshold. If growing mode energy is thermalized at a core/atmosphere boundary, rapid Helium burning and inward-travelling p-waves may result in core detonation. Thermalization at a boundary in the core can also result in detonation. If energy can be efficiently transferred from the orbit to modes as the WD binary passes through resonances, the WD merger timescale will be shortened by Myr-Gyr compared to expected timescales from GW-emission alone and GW detectors will observe deviations from predicted chirp profiles in resolved WD binaries. Future work in this area should focus on whether tidal locking in WD binaries is naturally driven towards low-order mode frequencies.
\end{abstract}


\section{Introduction}
White dwarfs (WD) are believed to be responsible for Type Ia supernovae (SNe); however the nature of the detonation and the (presumed) binary partner are still unknown. In order for WD merger rates derived from binary population synthesis (BPS) models to account for the observed Type Ia supernova rate, all WD binaries that merge through the emission of gravitational radiation within a Hubble time, including those with total binary mass below the limiting Chandrasekhar mass, must detonate \citep[see review by][]{Maoz14}. Most WD observed in our Galaxy are narrowly distributed around $M_{\rm WD}\approx 0.6M_{\odot}$ \cite{Kepler07}. Observations of Type Ia SNe imply an underlying $^{56}\rm{Ni}$ mass of $0.2-0.8M_{\odot}$ powering the SN and a total ejected mass spanning $0.8-1.5M_{\odot}$, albeit with some additional uncertainties  \citep{Scalzo14}. The detonation of substantially sub-Chandrasekhar mass WD would therefore be an elegant solution to the problem of the progenitors of SNe \citep[e.g.][]{vanK10}. In order to detonate a Carbon-rich WD, the runaway process of Carbon burning must occur \citep[e.g.][]{Martin06}. This need only happen in a small fraction of the WD mass \citep[e.g.][]{Rosswog09}. Runaway temperatures can be reached via core compression in a WD merger \citep{Pakmor12}, however the internal temperature can be driven up without core compression. A WD-compact partner (WD-CP) binary radiates energy from its orbital motion, gradually shrinking via gravitational wave (GW) emission. However some of the orbital energy may be converted to mechanical heating of the WD, rather than GW seen by a distant observer. This happens when the frequency of the quadrupolar tidal forcing experienced by the WD matches the frequencies of quadrupolar f-, p- and g-modes in the WD, and their harmonics, such that resonance occurs \cite[see e.g.][]{Rathore05,FulLai11}. In principle, if enough energy can be deposited into a massive mode during resonance, and that energy is thermalized rapidly, without much orbital backreaction or WD structural change,  unstable Carbon fusion and WD detonation could result. The goal of this paper is to investigate whether , and under what circumstances, a single WD in a binary can be driven to detonation via tidal resonance. Our conclusions depend on the length of time spent by the WD at resonance. Therefore whether WD binaries are naturally driven to tidal lock at frequencies which match low-order modes in WD is an important open question (and beyond the scope of this work). In this paper, in \S\ref{sec:consen} we consider sources and sinks of energy during WD binary inspiral. In \S\ref{sec:therm} we compare cooling of WDs via thermal processes and GW-emission. In \S\ref{sec:incr}, we discuss the energy deposited in a resonant mode, drawing a distinction between saturated and unsaturated driving and in \S\ref{sec:fusion} we discuss how Carbon-fusion might occur due to mode driving. In \S\ref{sec:trunc} we point out the several processes that can truncate resonance and limit the growth in thermal energy of nuclei in resonant modes. In \S\ref{sec:deposited} we discuss where the mode energy is likely to be deposited and in \S\ref{sec:obs} we point out the observational consequences of such thermal heating, both electromagnetic and GW, and including likely supernovae and novae.

\section{Energy conservation with modes}
\label{sec:consen}
A WD-CP binary shrinks over time as it loses orbital energy due to gravitational wave (GW) emission. Some of the orbital energy can be  transferred by tidal forcing into WD quadrupolar oscillation modes at resonant frequencies rather than GWs. Conservation of energy requires 
\begin{equation}
\dot{E}_{\rm orb}+\dot{E}_{\rm spin}= \dot{E}_{\rm GW} + \sum_{\rm j, WD} \dot{E}_{j}+ \sum_{\rm j, CP} \dot{E}_{j}
\end{equation}
where $\dot{E}$ denotes the time derivative of energy and $E_{\rm orb}=-GM_{\ast}M_{\rm c}/2a_{b}$ is the binary binding energy, $E_{\rm spin}$ is the total binary spin energy, $E_{\rm GW}$ is the energy carried away by GWs and $E_{j}$ is the energy deposited into mode $j$ of the WD or CP,  which may be detectable electromagnetically \citep[e.g.][]{Rathore05,McK14}. If spin and the transfer of orbital energy into mode resonances are ignored then $\dot{E}_{\rm orb}=\dot{E}_{\rm GW}$ which leads to binary shrinking solely due to gravitational wave emission on the well-known timescale \citep{Peters64}
\begin{equation}
t_{\rm GW} \approx \frac{5}{128} \frac{c^{5}}{G^{3}} \frac{a_{\rm b}^{4}}{M_{\rm b}^{2} \mu_{\rm b}} (1-e_{\rm b}^{2})^{7/2}
\label{eq:tgw}
\end{equation}
where $M_{\rm b}=M_{\ast}+M_{\rm c}$ is the binary mass, $\mu_{\rm b}=M_{\ast}M_{\rm c}/M_{\rm b}$ is the binary reduced mass and $(a_{\rm b},e_{\rm b})$ are the initial binary semi-major axis and eccentricity respectively. For a small change $\delta a_{b} \ll a_{b}$ in semi-major axis due to GW emission alone, the orbital energy changes by $(1/2)E_{\ast}(R_{\ast}/a_{b})(\delta a_{b}/a_{b})$ over a timescale $\delta t_{\rm GW} \approx t_{\rm GW} (4 \delta a_{b}/a_{b})$ and ignoring terms of order $\delta a_{b}^{2}$.

\subsection{Forcing frequencies}
As the WD binary shrinks, periodic tidal forcing generated by the binary increases in amplitude and can resonate with the jth quadrupolar $\ell =2$ eigenmode of the WD at angular frequency $\omega_{j}$ (with the usual associated mode numbers $n, \ell, m$). A tidal torque $T^{\ell m}_{n}$ acts on the resonant mode, where $n$ is the mode radial order (number of nodes) and $|m| \leq \ell$, but we shall restrict ourselves to $\ell,|m|=2$ \citep{Rathore05}. Harmonics of the tidal potential can drive WD modes at forcing frequencies 
\begin{equation}
\omega_{\rm F}=n\Omega_{\rm orb} -m\omega_{\rm spin}
\label{eq:forcingfreq}
\end{equation}
where $\Omega_{\rm orb}=\sqrt{\rm{GM_{b}}/a_{b}^{3}}$ is the WD orbital frequency and $\omega_{\rm spin}=f_{\rm spin} \omega_{\ast}$ is the WD spin frequency, with $f_{\rm spin}=[-1,+1]$ the fraction of the WD break-up spin frequency $\omega_{\ast}=\sqrt{\rm{GM_{\ast}}/R_{\ast}^{3}}$. The most promising case for sustained tidal forcing of a mode, so that the energy deposited is a maximum, occurs during tidal locking, when $\dot{\Omega}_{\rm orb} \approx \dot{\omega}_{\rm spin}$ so that $\omega_{F}$ remains approximately constant even as the orbit changes \citep{ws99}. Throughout this paper, we emphasize that mode energy is maximized if tidal locking in WD binaries is naturally driven to match low-order oscillation frequencies of a WD so that the mode is saturated. Under conditions of saturated driving, WD detonation could occur. 

For an equal-mass binary, ($M_{b}=2M_{\ast}$), eqn.~(\ref{eq:forcingfreq}) implies
\begin{equation}
a_{b}=2^{1/3}n^{2/3}\left( \frac{\omega_{F}}{k \omega_{\ast}}+2f_{\rm spin}\right)^{-2/3}R_{\ast}
\label{eq:abin}
\end{equation}
where $k=1,2,3,\ldots$ allows for tidal forcing at the $k$th harmonic of a mode frequency. For tidal resonances with modes involving oscillations of a moderate fraction of the WD mass, we require low radial order ($n \leq 10$). The condition for $n$ is
\begin{equation}
n=\frac{1}{\sqrt{2}}\left( \frac{\omega_{\rm F}}{k\omega_{\ast}}+2f_{\rm spin}\right)\left( \frac{a_{b}}{R_{\ast}}\right)^{3/2}
\label{eq:n}
\end{equation}
From \citep{Burkart13} we estimate  $\omega_{\rm F} \approx 0.1\omega_{\ast}-1 \omega_{\ast}$ for low-order g-modes, $\omega_{\rm F} \approx 1.5\omega_{\ast}$ for a f-mode and $\omega_{\rm F} \approx 2\omega_{\ast}-10\omega_{\ast}$ for low-order p-modes. In order to force a WD at these low-order ($n \leq 10$) quadrupolar mode frequencies, $a_{b}$ is required to be small in eqn.~(\ref{eq:abin}) and only certain combinations of harmonics and spins are allowed in eqn.~(\ref{eq:n}). For example if a p-mode of $4\omega_{\ast}$ is forced every 5th oscillation ($k=5$) and $f_{\rm spin}=+0.1(-0.5)$, then from eqn.~(\ref{eq:abin}), $a_{b} \leq 6(17.5)R_{\ast}$ for $n \leq 10$. Interestingly, retrograde spin $f_{\rm spin}<0$ allows forcing of harmonics of low-order ($n\leq 10$) p-modes at modest values of $a_{b}/R_{\ast}$ in eqn.~(\ref{eq:n}).  Field WDs generally rotate slowly \citep{Berger05}, but occasional rapid rotators may be due to mergers \citep{Ferrario05}. WD spin can significantly modify g-modes when $\Omega_{\rm orb} <\omega_{\rm spin}$ and $\omega_{\rm spin} \gg \omega_{F}$ \citep{FulLai14}. However, here we ignore spin effects such as g-mode modification or precession of misaligned spins. 

Resonance due to e.g. tidal locking will not be exact, i.e. the forcing frequency will be driven to a value away from exact resonance \citep{FulLai11}, leading to a resonant amplitude lower than peak by a modest numerical factor \citep{Burkart13}, and see also \S\ref{sec:heatcool} below. There is also still substantial uncertainty surrounding values of many of the physical parameters of binary white dwarf systems relevant to the discussion in this paper e.g. the exact frequency at which resonant lock will occur, the mass in particular eigen-modes, the likely frequencies of eigenmodes as well as the composition of white dwarfs close to merger. In the following discussion therefore, we approximate many of these and dependent quantities, using the symbol $\approx$ in that context, unless otherwise stated.
We will also for convenience use a model of a $\rm{M_{\ast}}=0.6M_{\odot}$ WD, since this is moderately well-understood \citep{Burkart13}, even though this is at the low mass end of the likely ejected SN mass and $^{56}\rm{Ni}$ mass.

\subsection{Mode heating \& cooling}
\label{sec:heatcool}
The equation of motion for the resonating jth mode of mass $M_{j}$ is:
\begin{equation}
\ddot{x_{j}} + \frac{\dot{x_{j}}}{\tau_{\rm d}}+\omega_{j}^{2}x_{j} =\frac{F_{j}}{M_{j}}
\label{eq:motion}
\end{equation}
where $x_{j}$ is the displacement of the $j$th mode, $F_{j}$ is the overlap integral beween the mode and the driver and $\tau_{d}=1/\Delta \omega_{j}$ is the decay time of the mode. For a constant (tidal) forcing frequency $\rm{F}=|F|e^{i \omega_{\rm F} t}$, the maximum steady-state displacement $x_{\rm max}$ is 
\begin{equation}
x_{\rm max} =\frac{|F|}{\sqrt{(\omega_{F}^{2}-\omega_{j}^{2})^{2}+(\omega_{F}/\tau_{d})^{2}}} 
\label{eq:xmax}
\end{equation}
which yields $x_{\rm max}=|F| \tau_{d}/\omega_{j}$ in the resonant limit ($\omega_{j} = \omega_{\rm F}$). If the $k$th harmonic of a resonant mode is being forced, the associated Fourier amplitude of that harmonic is lower than the $k=1$ mode amplitude by a factor $1/k$. In this case, the maximum displacement during oscillation will be reduced by a factor of $1/k$. 

Even at direct resonance with a mode, the maximum value of displacement in eqn.~(\ref{eq:xmax}) is unlikely to be reached in reality, since the forcing frequency will tend to lock slightly off-resonance \citep[e.g.][]{FulLai11,Burkart14}. Whether a tidal lock can occur and whether the resulting fixed frequency point is stable, unstable or chaotic, depends on the drift rate of the driver, the resonance width $1/\tau_{d}$, the rate of change of spin and the mass in the mode \citep{Burkart14}. For convenience we assume a forcing frequency $\omega_{F} = \omega_{j} + \delta \omega$, displaced from exact resonance by $\delta \omega =1/(r2\tau_{d})$, a multiple $r=1,2,3,\ldots$ of the half-width half maximum (HWHM) of the resonance peak. In this case, by expanding eqn.~({\ref{eq:xmax}) in a Taylor series and ignoring terms of order $(1/\tau_{d}^{3})$ and higher, the actual maximum displacement is reduced by a factor $1/\sqrt{2r}$. Ignoring the transient part of the general solution of eqn.~(\ref{eq:motion}), the real part of the solution to eqn.~(\ref{eq:motion}) includes the relative phase of the driver and the oscillator, as well as the phase lag of the forced oscillations. The effective forcing time ($t$) experienced by a mode is $t=[\tau_{d},1/\sqrt{\dot{\omega_{\rm F}}}]$ in the [saturated, unsaturated] case (see \citep{McK14} for an extensive discussion of this important distinction). The mode is saturated if the actual forcing time $t_{\rm F} >\tau_{d}$ and unsaturated if $t_{\rm F}<\tau_{d}$. Whether a mode is saturated or unsaturated makes a significant difference to the energy deposited into the mode. We can express the change in mode energy in terms of heating work ($W_{j}$) done on the mode and the cooling ($Q_{j}$) of the mode, namely
\begin{equation}
\dot{E}_{j}=\dot{W}_{j} - \dot{Q}_{j}.
\label{eq:1stlaw}
\end{equation}
Unlike main sequence stars for whom cooling via mode coupling dominates \citep{McK14}, a WD mode can cool efficiently via GW emission, mode coupling and neutrino emission \citep{Osaki73,Chandra91}. We ignore neutrino emission which is not important for the dissipation of oscillations in WDs \citep{Osaki73}, although it is important in releasing a significant fraction of the energy from the exothermic Carbon-fusion reaction. Work done on the mode that emerges as cooling via GW emission ($Q_{\rm GW}$) escapes the WD and will not increase the temperature of nuclei in the mode ($Q_{\rm kT}$). Thus for a given mode $j$, we can write the decay time of the mode as the sum of the GW and thermal cooling times $\tau_{\rm d} \approx \tau_{\rm j, GW}+\tau_{\rm j, kT}$, where $\tau_{\rm j,GW(kT)}$ are the GW(thermal) cooling timescales for the $j$th mode.

\subsection{Ratio of thermal to non-thermal cooling}
We write the average rate of mode cooling in terms of the non-thermal (GW) and thermal components as
\begin{equation}
\dot{Q}_{j}=\frac{Q_{\rm j, GW}}{\tau_{\rm j, GW}}+\frac{Q_{\rm j,kT}}{\tau_{\rm j,kT}}
\end{equation}
where
\begin{equation}
\tau_{\rm j, GW} \approx \left(\frac{5\rm{c}^{5}}{\rm{G}}\right) \left(\frac{1}{E_{j}\omega_{j}^{2}}\right) .
\label{eq:gwcool}
\end{equation}
Given that a quadrupolar mode (such as the ones considered here) can cool via GW emission as well as thermal dissipation, we find it useful to characterise any stellar/post-stellar mode generally in terms of the ratio of thermal to non-thermal cooling of a mode ($\mu_{j}$) as  
\begin{equation}
\mu_{j}=\frac{Q_{\rm j, kT}}{Q_{\rm j, GW}} \approx \frac{\tau_{\rm j, GW}}{\tau_{\rm j,kT}}
\label{eq:mu}
\end{equation}
which can be conveniently parameterized as
\begin{eqnarray}
\mu_{j} &=& 2.4 \times 10^{5} \left(\frac{M_{\ast}}{M_{\odot}}\right)^{-5/2} \left(\frac{R_{\ast}}{R_{\odot}}\right)^{5/2} \left(\frac{E_{j}}{E_{\ast}}\right)^{-1} \nonumber \\
&\times& \left(\frac{\omega_{j}}{\omega_{\ast}}\right)^{-1} \left(\frac{q_{j}}{10^{9}}\right)^{-1}
\label{eq:muj}
\end{eqnarray}
where $q_{j}=\omega_{j}\tau_{\rm kT}/\pi$ is the mode quality factor and $E_{\ast}=\rm{GM_{\ast}^{2}/R_{\ast}}$ in the WD binding energy. $\mu_{j}$ is smallest (i.e. GW emission is important) for excited quadrupolar modes in dense objects like WDs given a large mode energy, large decay time/narrow resonance width ($q_{j}>10^{9}$) at frequencies $\omega_{j} \approx \omega_{\ast}$. Note that we ignore the presence of an additional thermal energy source from exothermic Carbon-fusion reactions; this will only add to thermal energy in the mode, modulo losses due to neutrino emission. To first approximation, below the C-fusion threshold, most of the exothermic energy released due to Carbon-fusion will dissappear via free-streaming neutrinos. However, neutrinos will not carry away more than the Carbon-fusion energy.

The thermalization of mode energy dominates GW emission when $\mu_{j}\gg 1$, i.e. $\tau_{\rm j,kT} \ll \tau_{\rm j,GW}$. Physically, in the limit of large numbers of daughter modes, energy dissipated to daughter modes is effectively 'thermalized' to the nuclei involved in the oscillation. Mode coupling tends to occur near boundaries where the turning points of multiple nodes overlap, so the thermal heating will tend to dissipate at these boundaries \citep{kumargoodman96} and see also \S\ref{sec:deposited} below. Thus, nuclear heating will be localized at these boundaries if $\mu_{\rm j} \gg 1$ in a dissipation time $\tau_{\rm j,kT}$. 

\section{Thermalization vs Escape}
\label{sec:therm}
Generating a high local temperature for nuclei at mode turning points requires a short mode decay time for two reasons: (i) to keep GW losses small and (ii) to minimize radiative, conductive and convective thermal losses. \cite{kumargoodman96} show that $\tau_{\rm j,kT}$ decreases as a function of increasing mode energy, although for small $E_{j}/E_{\ast}$, $\tau_{\rm j,kT}$ is approximately constant for a given mode. At modest $E_{j}/E_{\ast}$, low order g-modes decay efficiently via mode coupling with an energy dependence \citep{kumargoodman96} 
\begin{equation}
\tau_{\rm j,kT} \approx 4\rm{\pi} \omega_{j}^{-1} \kappa^{-1}E_{j}^{-1/2} \approx \frac{4\rm{\pi}}{\omega_{j}}\left(\frac{E_{\ast}}{E_{j}}\right)^{1/2}
\label{eq:gmodedecay}
\end{equation} 
where the mode coupling coefficient is $\kappa \approx\rm{ G^{-1/2}R_{\ast}^{1/2}M_{\ast}^{-1}} \approx E_{\ast}^{-1/2}$ where $E_{\ast}$  is the WD binding energy. For ease of calculation, throughout this paper (unless otherwise stated) we shall use a conservative value of $\rm{M}_{\rm WD}=0.6M_{\odot}$ and $\rm{R}_{\rm WD}=0.013R_{\odot}$, which is at the low mass end of the range we expect \citep{Scalzo14}, but which corresponds to a relatively well-understood model \citep[e.g.][]{Burkart13}. We obtain $\mu_{\rm j} \approx 4 \times 10^{10}(E_{j}/E_{\ast})^{-1/2}$ for $\omega_{g}=0.1\omega_{\ast}$. So, excited quadrupolar g-modes in WDs \emph{always} cool via mode coupling rather than via GW emission,  and mode nuclei will be heated over timescale $\tau_{\rm j,kT}$.

Also at modest $E_{j}/E_{\ast}$, excited quadrupolar f-modes or low-order p-modes couple to $\ell=0,2,4$ p-modes leading to decay timescales of \citep{kumargoodman96}
\begin{equation}
\tau_{\rm j,kT} \approx \frac{2}{E_{j} \sum( \kappa^{2}/\tau_{p})} \approx 2 \frac{\overline{\tau}_{p}}{N_{p}} \frac{E_{\ast}}{E_{j}}
\label{eq:fmodedecay}
\end{equation}
where the sum is over the number ($N_{p}$) of the low radial order $\ell=0,2,4$ p-modes with typical decay time $\overline{\tau}_{p}$ that couple to the quadrupolar f-mode or low-order p-modes. Assuming some fraction ($f_{\rm GW} \approx 0.5$) of the daughter modes cool via GW emission, we find
\begin{eqnarray}
\mu_{\rm j} &=& 3 \times 10^{9} \left(\frac{M_{\ast}}{M_{\odot}}\right)^{-3}\left(\frac{R_{\ast}}{R_{\odot}}\right)^{4} \left(\frac{\omega_{j}}{\omega_{\ast}}\right)^{-2} \nonumber \\
&\times& \left(\frac{f_{\rm GW}}{0.5}\right) \left(\frac{N_{p}}{10^{2}}\right) \left(\frac{\overline{\tau}_{p}}{10^{10}\rm{s}}\right)^{-1}
\end{eqnarray}
or $\mu_{\rm j} = 2 (\overline{\tau}_{p}/10^{12})$ s for $M_{\rm WD}=0.6M_{\odot}$, $R_{\rm WD}=0.013R_{\odot}$ and $\omega_{\rm F}=1.5\omega_{\ast}$. So, excited f-mode and low-order p-modes cool via mode coupling rather than GW emission only when $\overline{\tau}_{p} \lesssim 10^{12}$ s (and we expect this condition will be satisfied for most such modes). 

\section{Increasing Energy Deposition}
\label{sec:incr}
The rate of work done per unit mass in the steady-state (saturated) case is $\dot{W}_{s} \approx \langle F^{2}\rangle \tau_{\rm j,d}$ and in the unsaturated case $\dot{W}_{u} \approx \dot{W}_{s}(t/\tau_{\rm j,d}) \approx \langle F^{2}\rangle t$ \citep{McK14}. The resonant tidal torque on the WD ($T=R_{\ast} |F| \rm{sin} \theta$)  yields a cycle averaged $\langle |F| \rangle \approx 2T/R_{\ast}$. During resonant tidal locking ($\dot{\omega}_{F} \approx 0$ and $\omega_{F} \approx \omega_{j}$) where $t$ becomes large \citep{ws99,Burkart14}, saturation becomes possible and we can use expressions for the torque \citep{wein12,Burkart13} to write
\begin{equation}
\langle |F| \rangle \approx \left(\frac{12\pi}{5}\right) \left(\frac{E_{\ast}}{R_{\ast}}\right) \epsilon^{2} \chi_{j}^{2} k^{-1} \left(\frac{\omega_{j}/\tau_{d}}{\delta \omega^{2} +(1/\tau_{d}^{2})}\right)
\label{eq:forcing}
\end{equation}
where $k$ is the harmonic of the mode and $\epsilon$ is the tidal factor
\begin{equation}
\epsilon=\left(\frac{M_{c}}{M_{\ast}}\right) \left(\frac{R_{\ast}}{a_{b}}\right)^{3},
\end{equation}
with $\delta \omega=\omega_{j}-\omega_{\rm F}$ the detuning frequency,  and $\chi_{j}$ is the overlap integral between the driving quadrupolar tidal force and the quadrupolar oscillation mode. Here we assume that $\chi_{j}$ for the $k$th harmonic is unchanged, but the amplitude of the forcing is diminished by factor $1/k$. As the resonant driving frequency drifts across resonance ($\delta \omega \approx 0$), a mode driven for time $t$ acquires total energy $E_{tot}=M_{j}|F/M_{j}|^{2}t^{2}/2$ \citep{McK14}, where $M_{j}=\chi_{j}M_{\ast}$ is the mass in the jth mode. Assuming the resonance is not exact and that $\delta \omega 1/(r2\tau_{d})$ is some multiple ($r=1,2,3,\ldots$) of the HWHM of the resonance (see discussion in \S\ref{sec:heatcool} above) we find 
\begin{equation}
E_{tot} \approx \left(\frac{6\pi^{2}}{5}\right)^{2} E_{\ast} \omega_{\ast}^{2} q_{j}^{2}\epsilon^{4} k^{-2} r^{-1}\chi_{j}^{3} t^{2}
\label{eq:emu1}
\end{equation}
where $q_{j}=\omega_{j} \tau_{j,kT}/\pi$ is the quality factor for the mode, and the mode is driven for time $t=t_{\rm F}(\tau_{d})$ in the unsaturated (saturated) case yielding $E_{tot} \propto \tau_{d}^{4}$ (saturated) and $E_{tot} \propto \tau_{d}^{4}(t_{F}/\tau_{d})^{2}$ (unsaturated). The energy transfer can be sensitive to $e_{b}$  \citep{Rathore05}, but here we shall ignore this complication. 

\subsection{Saturated modes}
\label{sec:sat}
During tidal locking, $\dot{\Omega}_{\rm orb} \approx \dot{\omega}_{\rm spin}$, so that the tidal forcing frequency remains approximately constant even as the orbit shrinks \citep{ws99}. Thus, if the locked tidal forcing frequency $\omega_{F}$ matches the frequency of a low-order WD mode (or its harmonic), then the mode can be resonantly driven for as long as tidal-locking lasts. If tidal-locking lasts longer than the mode decay time ($\tau_{d}$), then the mode is saturated. Substituting the equal-mass $M_{c}=M_{\ast}=0.6M_{\odot}$ Carbon-Oxygen WD binary model (CO6) of \cite{Burkart13} into eqn.~(\ref{eq:emu1}), we find the total energy deposited into a saturated low-order g-mode is
\begin{eqnarray}
E_{tot}&\approx &10^{-5}E_{\ast}\left(\frac{\omega_{j}/\omega_{\ast}}{0.1}\right)^{-2/3}\left(\frac{\chi_{\j}}{10^{-3}}\right) \nonumber \\
&\times&\left( \frac{r}{1}\right)^{-1} \left(\frac{M_{c}/M_{\ast}}{1}\right)^{4/3}\left(\frac{a_{b} /R_{\ast}}{10}\right)^{-4}
\label{eq:enparam_g}
\end{eqnarray}
where we have used $t=\tau_{\rm j, kT}$ from eqn.~(\ref{eq:gmodedecay}) and $\chi_{j}\approx 27 (\omega_{j}/\omega_{\ast})^{3.7} \approx 10^{-3}$ for $\ell=2$ g-modes with $n<10$ from \cite{Burkart13}. For a saturated low-order p-mode, forced at the $k$th harmonic, applying $t=\tau_{\rm j,kT}$ from eqn.~(\ref{eq:fmodedecay}) we find
\begin{eqnarray}
E_{tot}&\approx&0.1 E_{\ast}\left(\frac{\omega_{\ast}}{0.3\rm{s}^{-1}}\right)^{4/5} \left(\frac{\omega_{j}/\omega_{\ast}}{4}\right)^{2/5}\left( \frac{k}{1}\right)^{-2}\nonumber \\
&\times&\left(\frac{M_{c}/M_{\ast}}{1}\right)^{4/5}\left(\frac{a_{b}/R_{\ast}}{10}\right)^{-12/5} \left(\frac{\chi_{j}}{10^{-3}}\right)^{3/5}\nonumber \\
&\times&\left(\frac{\overline{\tau}_{p}}{10^{6}s}\right)^{4/5}\left(\frac{f_{\rm GW}}{0.5}\right)^{-4/5}\left(\frac{N_{p}}{10^{2}}\right)^{-4/5} \left( \frac{r}{1}\right)^{-1}
\label{eq:enparam_fp}
\end{eqnarray}
While the energy $E_{tot}$ in eqns.~({\ref{eq:enparam_g}) and ~(\ref{eq:enparam_fp}) is in principle available for deposition in the mode, multiple channels for truncating resonance exist (see \S\ref{sec:trunc} below) and one or more of these channels is likely to operate before energy in the mode reaches the values above. Nevertheless, Fig.~\ref{fig:freq} shows $E_{tot}$ as upper-limits to the total energy deposited in saturated low-order g- and (harmonics of) p-modes as a function of $a_{b}(R_{\ast})$. In both cases we assume $r=1$ (resonance-lock at a detuning frequency of $1/2\tau_{d}$) and $k=1$ to illustrate the maximum possible energy available. For energies deposited into realistic p-modes forced at the $k$th harmonic, divide the value of $E_{tot}(E_{\ast})$ given by the solid-lines by a factor $k^{2}$. Red, horizontal dashed lines show the mode energy required for an average temperature of $6 \times 10^{8}$K for Carbon nuclei in the mode for $\chi_{j}=10^{-3}$  (lower), $\chi_{j}=10^{-5}$ (upper). The red, horizontal dot-dashed line shows the mode energy required for an average termperature of $10^{8}$K for Carbon nuclei in a mode for $\chi_{j}=10^{-3}$, at which temperature atmospheric Helium ignition may occur (see \S\ref{sec:deposited} below). The black dashed line in Fig.~\ref{fig:freq} denotes the theoretical upper limit to mode energy in an equal mass binary, namely $E_{\rm orb}=-GM_{\ast}^{2}/2a_{b} \approx -E_{\ast}(R_{\ast}/2a_{b})$. The top axis of Fig.~\ref{fig:freq} indicates times to merge via GW emission alone (in Myrs) at binary separations of $a_{b}=10,20,50,100R_{\ast}$. Mode excitation to detonation is only possible for small $a_{b}$. From Fig.~\ref{fig:freq}, saturated, low-order g-modes with $\chi_{j} \approx 10^{-3}$ (at $a_{b}<5R_{\ast}$) and p-modes with $\chi_{j}>10^{-5}$ ( driven at harmonic $k$) could in principle absorb enough orbital energy so that Carbon-fusion could occur. Modes with small overlap integrals ($\chi_{j}<10^{-3}$ for g-modes and $\chi_{j}<10^{-5}$ for p-modes) cannot be driven to runaway Carbon-fusion temperatures. 

\subsection{Unsaturated modes}
\label{sec:unsat}
Away from resonance lock, at small $a_{\rm b}$, modes will tend to be unsaturated. Unsaturated resonant modes are important since their thermalization energy heats the WD allowing g-modes to exist and their backreaction on the orbit could perturb the orbit from $e_{b}=0$ (see below). Assuming $t_{F} \approx$ks (a few orbits at $a_{b} \approx 10R_{\ast}$), $E_{tot}$ for a low-order g-mode is reduced from the value in eqn.~(\ref{eq:enparam_g}) by a factor $10^{-6}(t_{F}/10^{3}\rm{s})^{2/3}(\tau_{d}/10^{12}\rm{s})^{-2/3}$ in the unsaturated case. Similarly $E_{tot}$ for a low-order p-mode is reduced from the value in eqn.~(\ref{eq:enparam_fp}) by a factor $0.06 (t_{F}/10^{3}\rm{s})^{2/5}(\tau_{d}/10^{6}\rm{s})^{-2/5}$ in the unsaturated case. The temperature of nuclei in unsaturated modes will be lower than in saturated modes by a factor $(t_{F}/\tau_{d})$ (see below). Unsaturated g-modes will never reach the Carbon-fusion threshold, but might reach the Helium-fusion threshold for modest forcings, or small departures from saturation. Unsaturated p-modes could reach both the C-fusion and He-fusion thresholds if the forcing time is relatively long ($t_{F}/\tau_{d}>0.1$). The f-modes ($n=0$) will generally be unsaturated since resonance lock ($\dot{\omega}_{F} \approx 0$) can only occur if spin remains constant while $\dot{a}_{\rm b}<0$. 

\begin{figure}
 \includegraphics[width=6.0cm,clip=true,angle=-90]{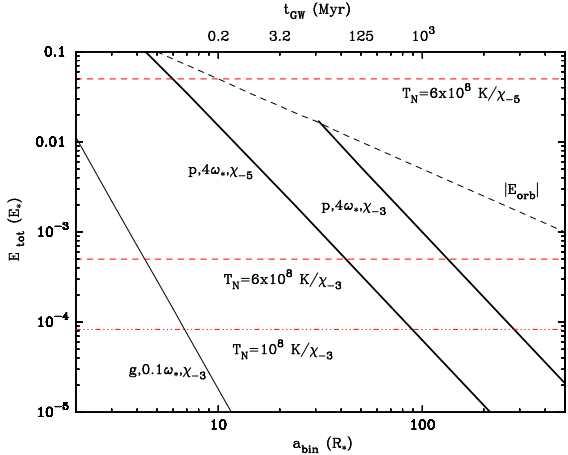}
\caption{\label{fig:freq} Average energy deposited in a saturated low-order mode as a fraction of orbital energy ($E_{\ast}$) versus binary semi-major axis $a_{b}(R_{\ast})$, for a low order p-mode ($4\omega_{\ast}$, thick lines) and a g-mode ($0.1\omega_{\ast}$, thin line),  with overlap integrals of $\chi_{j}=10^{-3},10^{-5}$  denoted by $\chi_{-3,-5}$ respectively. The p-mode energy (thick lines) corresponds to the $k=1$ upper limit. To find the actual limiting energy deposited, divide by $k^{2}$, for the $k$th harmonic of the p-mode being forced. Red horizontal dashed lines correspond to the Carbon-fusion instability temperature for $\chi_{-3}$ (lower) and $\chi_{-5}$ (upper). Red horizontal dash-dot line corresponds to a temperature of $10^{8}$K for $\chi_{-3}$ at which He-burning may occur in a thin atmosphere or thin envelope if mode energy is thermalized there (see text). Black dashed line shows the orbital energy $|E_{\rm orb}|=E_{\ast}(R_{\ast}/2a_{b})$, the theoretical maximum energy available in the binary system. The tidal disruption radius is $a_{b} \approx 2R_{\ast}$ using $M_{\ast}/M_{c} \approx 1$ \citep{egg83}.
On the top horizontal axis are the merger timescales (in Myrs) calculated for GW emission alone for an equal mass $0.6M_{\odot}$ WD binary corresponding to binary separations of $a_{b}=10,20,50,100R_{\ast}$.  So e.g. if detonation occurs at $a_{b}=50R_{\ast}$, it would be $125$Myr before the merger predicted by GW emission alone.
}
\end{figure}

\section{Igniting Carbon fusion}
\label{sec:fusion}
A WD mode consists of two parts: the nuclei and the electrons. The state of the nuclei depends on the Coulomb coupling parameter $\Gamma=(Ze)^{2}/a_{N}k_{\rm B}T$ or
\begin{equation}
\Gamma \approx 25 \left(\frac{Z}{6}\right)^{2}\left(\frac{A}{12}\right)^{-1/3}\left(\frac{T}{10^{7}\rm{K}}\right)^{-1}\left( \frac{\rho}{4 \times 10^{5} \rm{g cm^{-3}}}\right)^{1/3}
\end{equation}
where $a_{N}$ is the inter-nucleus spacing, $k_{\rm B}$ is the Boltzmann constant, $T$ is the temperature of the nuclei, $Z$ is the atomic number, $A$ is the atomic weight, $e$ the electron charge and $\rho$ is the (uniform) WD density \citep{kippen90}. If $\Gamma >180$, the nuclei form a solid crystal lattice and if $\Gamma \lesssim 1$, the nuclei behave as an ideal gas with heat capacity $C_{v}=3/2 n_{N} k_{\rm B}T_{N}$, so mode heating/cooling changes the (local) nuclear temperature and can lead to structural changes in the WD.

Total mode energy is $E_{tot,j}=E_{tot,N}+E_{tot,e} =N_{N}\overline{E}_{N} + N_{e} \overline{E}_{e}$
where $N_{N},N_{e}$ are the number of nuclei and electrons in the mode respectively and $\overline{E}_{N}(\overline{E}_{e})$ are the respective  average nucleus(electron) energy. Assuming equipartition of the thermalized mode energy among the charged particles, and $N_{e}=6N_{N}$ for pure Carbon, the total energy in the mode nuclei $E_{tot,N}=E_{tot,j}/7=3/2 k_{\rm B}T_{N}$ where $T_{N}$ is the average temperature of mode nuclei. Energy $E_{tot,e}=(6/7)E_{tot,j}=p_{e}^{2}/2m_{e} (p_{e}c)$ is added to the non-relativistic (relativistic) degenerate electrons in the mode where average electron momentum is $p_{e}\approx \hbar n_{e}^{-1/3}$ and $m_{e}$ is the electron mass. Increasing $E_{tot,e}$ alters the electron pressure, potentially truncating resonance (see \S\ref{sec:trunc} below).

Energy deposited into the mode during resonance lock over integrated forcing time $t_{\rm F}$ is thermalized among the Carbon nuclei on damping timescale $\tau_{d}$, which is so short we can neglect losses to conduction, convection and radiation \citep{Osaki73}. Heating of the nuclei in a mode yields 
\begin{equation}
\int^{t_{\rm F}}_{0}\frac{E_{\rm tot,j}}{7 \tau_{d}} dt \approx \frac{3}{2}N_{\rm N} k_{\rm B} T_{\rm N}
\end{equation}
so 
\begin{equation}
T_{\rm N} \approx \frac{2}{21}E_{\rm tot,j} \left(\frac{t_{F}}{\tau_{d}}\right) \frac{1}{k_{\rm B} N_{\rm N}}
\label{eq:temp}
\end{equation}
where $N_{\rm N}=\chi_{j}M_{\ast}/m_{\rm C12}$ is the number of nuclei in the mode, and $m_{\rm C12}$ is the mass of the Carbon 12 atom, assuming all the nuclei are Carbon 12. In the saturated case $t_{f} \approx \tau_{d}$ so $T_{\rm N} \approx (2/21)(E_{\rm tot}/k_{\rm B}\chi_{j})(m_{\rm C12}/M_{\ast})$. Thus, for a saturated mode $T_{\rm N} \approx 6 \times10^{8} \rm{K} (E_{\rm tot}/5 \times 10^{-4}E_{\ast}) (\chi_{j}/10^{-3})^{-1}$ shown by the red dashed horizontal lines in Fig.~\ref{fig:freq} for $\chi_{j}=10^{-3},10^{-5}$. From Fig.~\ref{fig:freq}, in principle C-fusion could occur among the nuclei in a g-mode with $\chi_{j} \approx 10^{-3}$ at $a_{b}\leq 5R_{\ast}$ and in a p-mode with $\chi_{j} \approx 10^{-5(-3)}$ at $a_{b}<3(60)R_{\ast}$ at small harmonics ($k<5$). 

\section{Truncating resonances}
\label{sec:trunc}
A WD is driven away from resonance either via structural change which alters mode frequency, or orbital back-reaction. An increase in WD electron energy decreases the WD radius by $\Delta R_{\ast}/R_{\ast}=6E_{j}/7E_{\ast}$ changing the frequency spectrum of WD modes by $\Delta \omega_{\ast}/\omega_{\ast} \approx (9/7)(E_{j}/E_{\ast})$. Resonance can only continue if the mode width $\Delta \omega_{j}<\Delta \omega_{\ast}$. In the absence of changes to $\Delta \omega_{j}$, we would expect (narrow) g-mode resonance to truncate after $E_{j}/E_{\ast}>10^{-12}$, and (wider) p-mode resonances to truncate after $E_{j}/E_{\ast}>10^{-6}$. However, from \S\ref{sec:therm}, modes widen with increasing energy deposition in the mode. Comparing $\Delta \omega_{\ast}$ to $\pi/\tau_{j,kT}$ from  eqn.~(\ref{eq:gmodedecay}), we find that structural changes in the WD will end resonance with g-modes for $E_{j}/E_{\ast}>4 \times 10^{-4}$ for $\omega_{j} \approx 0.1\omega_{\ast}$ and $E_{j}/E_{\ast}>10^{-2}$ for $\omega_{j} \approx 0.5\omega_{\ast}$. Thus, in principle low-order g-modes with $\omega_{j}>0.1\omega_{\ast}$ could maintain resonance up to ignition even as the WD radius decreases, by increasing $\Delta \omega_{j}$ as a function of mode energy. As long as a constant $\tau_{\rm j,kT}$ at low $E_{j}/E_{\ast}$ permits p-modes to be excited then resonance can persist. Comparing $\Delta \omega_{\ast}$ to $\pi/\tau_{j,kT}$ from eqn.~(\ref{eq:fmodedecay}), we find that low-order p-modes maintain resonance in the face of WD structural changes only if $N_{p}/\overline{\tau}_{p}>\omega_{\ast}$. Thus, only if $N_{p}/\overline{\tau}_{p}$ itself increases with $E_{j}/E_{\ast}$ (which we anticipate) can resonance be maintained for low-order p-modes in the face of WD structural changes.

Back-reaction is the transfer of energy from a resonant mode to the orbit when $\dot{\Omega}_{\rm orb} \geq \delta \omega_{j}$, the resonance width \citep{Rathore05}. Back-reaction can be ignored at modest separations and large energy transfer since $\dot{\Omega}_{\rm orb} \propto a_{b}^{-11/2}$ and $\tau_{d}^{-1} \propto E_{j}/E_{\ast}$ for p-modes, but where back-reaction occurs it  can reduce the energy transfer to the mode by up to an order of magnitude \citep{Rathore05}. In the absence of resonant locking, back-reaction can quickly (several orbital timescales) terminate unsaturated narrow resonances by increasing $a_{b}$ (if $e_{b} \approx 0$). Orbital back-reaction on unsaturated resonances is important since it can keep $e_{b}>0$ and maintain the tidal torque $T^{\ell m}_{n}$ on $m=2$ quadrupolar modes. If large mode energy gets transferred rapidly to the orbit, driving $e_{b} \gg 0$, tidal squeezing of the WD core by the CP at high orbital eccentricity will increase core density, potentially initiating C-fusion \citep{Rosswog09}. After GW emission of the back-reaction orbital energy, the WD can return to resonance. 

Absorption of sufficient resonant energy by a mode can also have the effect of \emph{accelerating} the shrinkage of a WD binary orbit compared to a binary shrinking by GW emission alone \citep{FulLai11}. In this case, the forcing frequency crosses the resonance width faster than expected and depending on the magnitude of $\dot{\omega}_{F}$, modes which would otherwise be saturated (forcing time is $t_{F}=\tau_{d}$) can become unsaturated (forcing time is now $t_{F}=1/\sqrt{\dot{\omega_{F}}}$). Thus, once $\tau_{d}>1/\sqrt{\dot{\omega_{F}}}$ then the resonance will be truncated and the mode will become unsaturated, with a resulting decrease in energy deposited in the mode (see \S\ref{sec:unsat} above). Further work is required to determine the expected total energy deposit in a real WDB.

Resonant amplification of an eigenmode is also likely to eventually lead to complicated non-linear hydrodynamical phenomena in the WD, including wave-breaking \citep[e.g.][]{FulLai12,Burkart13}. Such effects may in fact truncate the resonance itself, reducing the time spent at resonance and therefore the energy deposited in the mode, and the likely observational signature.

\section{Where is the mode energy deposited?}
\label{sec:deposited}
Orbital energy absorbed by a resonant mode will tend to transfer to daughter modes at locations in the WD where the nodes of multiple modes encounter each other \citep{kumargoodman96}. This will generally happen at boundaries within the WD at turning points of modes.
In the case of a uniform homogeneous sphere, with a thin atmosphere or envelope, the turning points of modes will lie only at the boundary of the WD with its atmosphere and therefore all of the mode energy will be transferred to daughter modes  and thermalized at the base of the atmosphere. If energy deposition into the mode from tides and energy leakage into an atmosphere or envelope equalize, a dynamical equilibrium would result in a steady burn rather than detonation-- possibly resulting in a nova rather than a supernova; we shall leave this calculation for future work. Here we are interested in the possibility of a more violent outcome. As the energy in the mode increases at a rate greater than energy loss, if the WD has a Helium atmosphere, mode energy will be thermalized at the base of the atmosphere and ultimately start Helium burning at temperatures ($\sim 10^{8}$K) well below C-fusion temperatures. Certain SNe, including Ca-rich SNe may be accounted for by the ignition of a thin He atmosphere around a low-mass Carbon core \citep{Waldman11}. If the Helium in a spherical atmospheric shell can ignite instantaneously, this induces a radially inward propagating p-wave that converges on the core and Carbon-ignition can result \citep{Livne90}, yielding a Type Ia SN before merger occurs. If Helium detonation is less symmetric, the Helium explosion and resulting core-compression may yield instead a sub-luminous Type Ia SN \citep{Bildsten07}, or a tidal nova \citep{FulLai12}. However, if the Helium can be ignited across a sufficiently large region, compressional waves induced by a sliding Helium detonation can trigger a detonation in the Carbon core \citep{MollWoosley13}. Indeed, even if the Helium detonation occurs in small patches in a thin He atmosphere, polluted with some C/O, a Type Ia SN can still result \citep{Shen14}. A very thin He atmosphere can suffice for detonation: mass $\approx 10^{-2}M_{\odot}$ in a $0.6M_{\odot}$ WD  and mass $5\times 10^{-3}M_{\odot}$ in a $1.0M_{\odot}$ WD \citep{Shen14}. Since most C/O WDs are surrounded by a thin He atmosphere \citep{Iben85}, the ignition of the base of the He atmosphere due to resonant mode thermalization is a very promising avenue for Type Ia SNe generation.

Heat dissipation and ignition can also occur at other WD locations. Essentially, once a boundary exists within the WD core, for example, if there is a small solid or liquid central WD core, mode turning points can dissipate heat at the boundary between the two phases of the central core. In this case, the mode energy will be deposited equally as heating at the base of the atmosphere and the boundary between the phases of the central WD core. Spherical or patchy ignition around the surface of a central solid or liquid core could act analagously to shell ignition in a He atmosphere \citep{Livne90,Shen14} and cause a inward travelling p-wave that ignites the core. Wave-breaking due to non-linear effects may truncate all of the above effects \citep{FulLai12} and may in fact result in the driving of a wind or greater radiation from the surface, although we ignore these possible limitations for the present work.

\section{Observational Implications}
\label{sec:obs}
If modes thermalize at the base of a thin He atmosphere, either in a single sheet or in a large number of patches (due to a large number of nodes), the result should be a spherical ignition of the Helium atmosphere and a spherically symmetric Type Ia SN \citep{Shen14}. If instead the number of nodes is small, the resulting He ignition will begin substantially asymmetrically and may result in edge-lit detonation of the C/O core, yielding an asymmetric Type Ia SN. Equally, if there is a slow, steady burn of an atmosphere or envelope, the result may well be a nova or radiative wind, and even if C-fusion commences the burn may not be explosive. We will need to understand which modes are most likely to resonate with tidal locking before we can make predictions about the asymmetry of the explosion.
  
WD detonation at binary separations $a_{b} <10^{2}R_{\ast}$ before merger imply Type Ia SNe can occur ($\approx$ Myr-Gyr) before mergers driven by GW-emission alone. Therefore binary population synthesis modelling should include this merger acceleration, in order to correctly model the observed Type Ia SN delay-time distribution \citep{Graur14}. Dissipation of unsaturated resonances within the WD (as well as tidal heating) will lead to WD rejuventation, so close binary WDs will be hotter than their age would suggest. Even if a WD starts off 'cold', it will quickly 'heat up' from thermalization of unsaturated mode resonances, decreasing $\Gamma$, allowing g-modes to exist. 

The energy for WD rejuvenation and potential detonation comes from $E_{\rm orb}$. Our model predicts the binary inspiral time will be changed by a factor $\dot{E}_{\rm orb}/E_{\rm kT}$, where $E_{\rm kT}$ is the total energy deposited in modes. Strong resonances will absorb orbital energy that would otherwise have emerged as GWs. Thus, shrinking WD binaries will skip orbits as they pass through resonances \citep{FulLai11}. 
\citet{Timpano06} show that in one year eLISA will resolve approximately $10^{4}$ binaries at S/N$>5-10$ of which around $300$ (mostly WD-WD) will be chirping. Extracting orbital energy and adding it to WD modes will tend to close $a_{b}$ faster than via GW emission alone. Therefore eLISA will detect an increase in $dh/dt$ over the $0$th order chirping profile during resonances \citep[see also][]{FulLai11}. During a resonance, $\dot{h}$ will increase over the value predicted by GW-emission alone by a factor $\tau_{d}/\Delta t_{\rm GW} \approx (1/4)(\tau_{d}/t_{\rm GW})(a_{b}/\delta a_{b})$, where $\delta a_{b} \ll a_{b}$ represents a small change in the binary semi-major axis. Depending on the spin-orbit interaction during resonance, orbital back-reaction from mode excitation that happens to result in an increase in $(a_{b},e_{b})$ for the WD binary (analagous to the dynamics of the Earth-Moon system), will yield a temporary \emph{decrease} in $h$ and $decrease$ in GW frequency, also detectable with eLISA. Of course, if WDs in binaries manage to detonate via resonant excitation before merger, then  eLISA will \emph{not} detect any chirping WDBs.

Note that if energy transfer to modes is efficient during resonance, regardless of whether a WD detonates, the merger timescale for WD binaries must decrease by a factor $\approx (N\overline{\tau_{d}}/\delta t_{\rm GW})$ as it passes through N resonances of mean-width $\overline{\tau_{d}}$, where $\delta t_{\rm GW}$ would be the timescale for shrinking $a_{b}$ by an amount $\delta a_{b}$ due to GW-emission alone. Therefore, regardless of the fate of the WD itself, the WD binary merger timescale must decrease from the value expected from GW-emission alone and this will increase the normalization of the delay-time-distribution predicted by binary population synthesis models.

\section{Conclusions}
We investigate the conditions under which WDs with sub-Chandrasekhar masses in binaries with compact objects (including other WDs) could resonantly detonate in Type Ia supernovae before merger. Requirements for detonation include: 1. resonant lock on a sufficiently massive  low-order quadrupolar mode or harmonic, 2. a modest fraction of orbital energy (up to $\approx 10^{-3} E_{\rm orb}$) is deposited into the mode over the forcing time without significant orbital backreaction or structural change \emph{and} 3. the mode energy is quickly thermalized among the Carbon nuclei in that mode. A very promising avenue for future work is whether WD binaries are naturally driven to tidal lock with low-order mode frequencies (or their harmonics), since this may be a channel for Type Ia SNe.

We find that ignition could in principle occur, Myr-Gyr before merger, during resonant tidal lock with a low-order $\ell=2$ g-mode with mass $\approx 10^{-3}M_{\ast}$ or with a harmonic of a low-order p-mode with mass $>10^{-5}M_{\ast}$ and in the limit of a large number of daughter modes. Depending on where the mode energy is thermalized and the efficiency of thermalization, ignition may first occur in a He atmosphere. In this case, a spherically symmetric ignition of the He atmosphere, or asymmetric ignition in a massive He atmosphere can lead to compressive p-wave shocking of the Carbon core and detonation. If there is a boundary in the CO interior of the WD, thermalization can also happen there. 

Resonances can be truncated by orbital backreaction or an excessive increase in electron degeneracy pressure. Sufficient orbital back-reaction from the excited mode may increase orbital eccentricity such that that thermalization of pericenter forcing generates Carbon-fusion. Sufficiently rapid extraction of orbital energy can allow the WD binary to pass through resonance fast enough that the resonant mode becomes unsaturated and the energy deposited in the mode drops significantly. 

All WDs in a close binary will be rejuvenated by thermalizing multiple resonances. GW detectors should detect a deviation from the predicted chirping strain amplitude ($h$) profile from WD binaries at frequencies matching low-order, high mass quadrupolar modes. Regardless of the fate of the WD, an efficient transfer of orbital energy to resonating modes will reduce WD binary merger timescales, which should be accounted for in binary population synthesis models in order to predict the observed delay-time-distribution of Type Ia SNe. Future work should focus on whether tidal locking in WD binaries is naturally driven towards low-order mode frequencies.

{\section{Acknowledgements.}} Thanks to the referee, Christopher Tout, for a report that greatly helped us clarify our arguments. We acknowledge very helpful and insightful criticisms from Jim Fuller and useful conversations with Or Graur, Zoltan Haiman, Saurabh Jha, Bence Kocsis, Cole Miller, David Tsang and Sterl Phinney. BM \& KESF are supported by NSF PAARE AST-1153335 and NSF PHY11-25915.\\
\\

\end{document}